# Errors and secret data in the Italian research assessment exercise. A comment to a reply

Alberto Baccini
Department of Economics and Statistics, University of Siena, Italy
Piazza San Francesco 7, 53100 Siena, alberto.baccini@unisi.it;

Giuseppe De Nicolao
Department of Electrical, Computer and Biomedical Engineering, University of Pavia, Italy

**ABSTRACT.** Italy adopted a performance-based system for funding universities that is centered on the results of a national research assessment exercise, realized by a governmental agency (ANVUR). ANVUR evaluated papers by using "a dual system of evaluation", that is by informed peer review or by bibliometrics. In view of validating that system, ANVUR performed an experiment for estimating the agreement between informed review and bibliometrics. (Ancaiani et al. 2015) presents the main results of the experiment. (Alberto Baccini and De Nicolao 2017) documented in a letter, among other critical issues, that the statistical analysis was not realized on a random sample of articles. A reply to the letter has been published by *Research Evaluation* (Benedetto et al. 2017). This note highlights that in the reply there are (1) errors in data, (2) problems with "representativeness" of the sample, (3) unverifiable claims about weights used for calculating kappas, (4) undisclosed averaging procedures; (5) a statement about "same protocol in all areas" contradicted by official reports. Last but not least: the data used by the authors continue to be undisclosed. A general warning concludes: many recently published papers use data originating from Italian research assessment exercise. These data are not accessible to the scientific community and consequently these papers are not reproducible. They can be hardly considered as containing sound evidence at least until authors or ANVUR disclose the data necessary for replication.

Since 2010 Italy adopted a performance based system for funding research and universities. The funding of universities is based on a national research assessment, called VQR, by and large inspired to the British experiences of RAE/REF. Italian research assessment was conducted by the Italian governmental agency for evaluation of universities and research (ANVUR). ANVUR adopted for the VQR "a dual system of evaluation" according to which each piece of work submitted was classified in one class of merit by informed peer review (IR) or by using bibliometric indicators. The results of evaluation reached by using these two different techniques were then combined for calculating overall scores for research fields, departments and universities. The basic idea behind this methodology is that IR and bibliometrics can be used interchangeably. This assessment design was adopted in the first edition of the VQR (2004-2010), and it was repeated also in the second edition (2011-2014).

In the first edition (hereafter VQR1), in view of validating the dual system of evaluation, ANVUR performed an experiment for estimating, for a large set of papers, the agreement between scores obtained by IR and by bibliometrics. Results of this experiment are central for the consistency of the whole research assessment exercise. The results of the experiment were originally published in the Appendix B of the ANVUR official report (ANVUR 2013). They were then widely disseminated in working papers and scholarly articles originating from or reproducing parts of the ANVUR reports (a list is available in Alberto Baccini and De Nicolao 2016a). This long list of papers probably aims to justify ex-post, through scholarly publications, the unprecedented methodology adopted by ANVUR. Overview of the critical remarks on the methodology can be found in (G. Abramo et al. 2013; Giovanni Abramo and D'Angelo 2015; Giovanni Abramo and D'Angelo 2017; Alberto Baccini and De Nicolao 2016a; A. Baccini 2016; Alberto Baccini and De Nicolao 2016b)

(Ancaiani et al. 2015) summarly describes and presents the main results of the experiment. (Alberto Baccini and De Nicolao 2017) documented in a letter (hereafter "our letter") critical issues and a major shortcoming in the design of the experiment. Indeed the analysis of statistical concordance was not conducted on a random sample of articles, but on a non-random subsample, obtained by excluding from the original random sample all the articles for which bibliometrics produced an uncertain classification.

A reply to our letter has been published by *Research Evaluation* as Benedetto et al. (2017) while the data used in the original article and in the reply continue to be undisclosed. That reply contains many unresolved issues that we highlight in this note under six heads.[1]

1. **Errors in data.**

Data presented in Benedetto et al. (2017) contain errors, possibly indicative of underlying problems in raw data.

- According to Table 1 of Benedetto et al., they are working on a population of 99,005 articles; in Table 2, the "Population" reduces to 86,998 articles (sum of the figures in the column labelled "Population"). The third column of Table 2 (labelled "%") contains therefore percentages that do not correspond to data of the population (Table 2, second column). Data of Table 2 are drawn from Appendix B of ANVUR report (ANVUR 2013 Tab. B3) that contains these same inconsistent data. Table 2 contains also a (minor) "factual error": the figure for population is 4,7583 instead of, probably, 47,583.

- Benedetto et al. confirm the main claim of our letter: they did not use the complete random sample for their investigation (9,199 articles), but a non-random subset of the random sample. In their reply they finally provided some (incomplete) information about this subset, but, unfortunately, the numbers are still inconsistent. Indeed, in Table 1 and Table 4 the subsample appears as having 7.598 observations; from Table 2 it is instead possible to calculate a subsample of 7.597 by summing-up evaluation classes from A to D (last column). This last figure coincides with the estimate contained in our letter, and also with ANVUR official data (ANVUR 2013, Appendice B). At the state of information it is impossible to know what the correct figure is.

---

[1] Authors communicated the contents of the six heads of this note to the editorial board of *Research Evaluation* in a letter of 6th June 2017. We agree also with editors' request to send that letter to the authors of the reply.

2. **Problems with "representativeness" of the sample**.

Benedetto et al. introduce an issue they call "representativeness" of the subset of the random sample. Contrary to what written by Benedetto et al., in our letter we did not make any claim on the (lack of) "representativeness of the subsample". Our claim is about the unknown biases induced in the results by selecting a non-random sub-set of data from a random sample. Apparently, Benedetto et al. confuse this question with that of "representativeness", and they did not address at all our claim in their reply.

In any case, their discussion of "representativeness" raises new issues:

- According to Benedetto et al., the "sample was stratified according to the scientific area of the author", and "representativeness" was considered at an Area level. This information is at odds with ANVUR official reports, according to which the sample was stratified at a "sub-area" level, the so-called sub-GEV (see page 1 of Appendix B of the VQR 2004-2010 Final Report: "The population was stratified based on the distribution within the sub-GEV's"). In particular, the stratification of the random sample was conducted by adopting oversampling for specific sub-areas. For economics and statistics, for example, the sample was stratified in four sub-areas, and for the sub Area Economic history the sample contained a 25% of the population of articles (ANVUR 2013; Bertocchi et al. 2015). It is therefore unclear why the "representativeness" of a sample stratified by sub-area is now evaluated at a different level of aggregation.

- Benedetto et al. wrote that their Table 1 was "not reported in the article for the sake of simplicity and lack of space". Note that Table 1 was not even reported in the thousands of pages of ANVUR on-line reports, where "lack of space" cannot be a matter. Possibly because the representativeness of the subsample was not computed at all. It is also a bit curious that in Ancaiani et al. (2015) the authors failed to disclose to *Research Evaluation*'s readers that they were working on a non-random sub-sample of 7,598 (or 7,597?) papers. This relevant information is very simple and would not have required a lot of space to be disclosed.[2]

- Benedetto et al. wrote that "Table 2 shows that ex-post distributions of bibliometric evaluation is pretty similar in the reference population and in the subsample". The reader is therefore induced to believe that Table 2, notwithstanding the column title reporting "sample", provides percentages for the 7,598 (or 7,597?)-item subsample. But this is not the case. Actually, Table 2 compares data about a 86,998-item population (where, as already noted, 12,007 papers have disappeared with respect to the total of Table 1), with data of the complete 9,199-item random sample (we found this, just by observing that the penultimate column sums up to 9,199, instead of 7,598 or 7,597, the dimension of the subsample). That table simply demonstrates that, as obviously expected, the distribution of a character (bibliometric evaluation) in the random sample is similar to the distribution of that same character in the population. Nothing is showed about the distributions in the population of the other variables (P1, P2, P) used in the analysis, and the joint distribution of F and P in the sample and, more importantly, in the subsample. Benedetto et al. cannot therefore assess the distortions possibly induced by the non-random selection of the subsample.

---

[2] This is a first problem relevant for the question of the integrity of research record as published in *Research Evaluation.* Indeed the journal did not publish a correction, and readers of Ancaiani et al. (2015) will continue to read that data refer to "a [representative] sample equal to 9,199 articles" while data refer to a non-random sub-set of 7,598 (or 7,597?) articles.

− Benedetto et al. obtained the subset of the random sample by discarding all papers classified as IR. Recall that a paper was classified as IR when its bibliometric evaluation was difficult, because it appeared in a journal with high Impact Factor but received few citations, or viceversa. We know from ANVUR reports how peers evaluated the papers classified as IR, and we know also how peers evaluated the whole random sample. It is therefore possible to compare, as in Table 1, the distributions of peer review scores in the random sample and in the set of excluded papers (IR).

**Table 1.** Comparison of the distributions of evaluations in the random sample and in the set of excluded papers (IR).

| Evaluation class attributed by peer review | Random sample | % | IR articles excluded by the random sample | % |
|---|---|---|---|---|
| A | 1,531 | 16.6 | 125 | 7.8 |
| B | 3,321 | 36.1 | 546 | 34.1 |
| C | 907 | 9.9 | 167 | 10.4 |
| D | 1,521 | 16.5 | 345 | 21.5 |
| IP | 1,919 | 20.9 | 419 | 26.2 |
| *Total* | 9,199 | 100.0 | 1,602 | 17.4 |

Source of data: ANVUR 2013. Appendix B. Table B.3

It is easy to see that peer-review evaluations of the papers excluded by the subsample are not distributed as in the random sample. In particular, in the excluded IR papers the share of D and IP papers is bigger than, and the share of A papers is smaller than in the random sample. The only way to rule out the possibility that the non-random selection of papers biases final results would be to perform a careful and thorough analysis of the raw data that continue to be not accessible to scholars.

− Benedetto et al. comment the last two column of their Table 1 by noticing "that small differences with respect to proportionality are due to the fact that the subsample of Area 13 coincides with the sample".
As a matter of fact, the observed differences have nothing to do with the subsample of Area 13. In Table 1, sample and subsample sizes are compared separately for each area: for example, in Area 1 the sample represents a 9.3% of the population while the subsample represents only a 6.5% of the population. Despite their claim, the size of the sample of Area 13 is completely uninfluential for the subsampling shares of other areas. More importantly, the differences with respect to proportionality are instead due to the non-random selection that discarded different shares of articles in each area, as illustrated in the next table. At the two extremes, for Mathematics and informatics the size underwent a 30.6% reduction against -12.0% for Civil Engineering and 0.0% for Economics and statistics where no article was discarded.

**Table 2.** Comparison of sample size, subsample size and number of discarded IR articles.

|  | Random sample | Subset of the random sample | Number of discarded IR articles | Percentage reduction (%) |
|---|---|---|---|---|
| Area 1 Mathematrics and informatics | 631 | 438 | 193 | **-30,6** |
| Area 2 Physics | 1,412 | 1,212 | 200 | -14,2 |
| Area 3 Chemistry | 927 | 778 | 149 | -16,1 |
| Area 4 Earth Sciences | 458 | 377 | 81 | -17,7 |
| Area 5 Biology | 1,310 | 1,058 | 252 | -19,2 |
| Area 6 Medicine | 1,984 | 1,602 | 382 | -19,3 |
| Area 7 Agricoltural and veterinary sciences | 532 | 425 | 107 | -20,1 |
| Area 8a Civil Engineering | 225 | 198 | 27 | -12,0 |
| Area 9 Industrial & information engineering | 1,130 | 919 | 211 | -18,7 |
| Area 13 Economics and statistics | 590 | 590 | 0 | **0,0** |
| *All areas* | *9,199* | *7,597* | *1,602* | *-17,4* |

Source of data: ANVUR 2013. Appendix B.

− Benedetto et al. report unfaithfully our statements when they write: "We hence conclude that, contrary to the claim of the authors of the letter, the evaluation of concordance has been performed on a sample that is fully representative of the original population of articles to be evaluated". As already anticipated, in our letter no claim is made on the (lack of) representativeness of the subsample for the simple reason that, as repeatedly pointed out, raw data are not available. It is also strange that Benedetto et al. prefer to show (inconclusive) statistics based on data not accessible to the scientific community, instead of answering the representativeness issue in the most obvious way, that is by disclosing raw data.

3. **Unverifiable claims about weights used for calculating kappas**.

In order to remedy inconsistencies pointed out in our letter, Benedetto et al. replace Table 2 of the original paper with Table 3 of the reply, where they modify three kappas.[3] Benedetto et al. claim that: (i) "differences are attributable to factual errors in the editing of the table" and (ii) "the same set of weights to compute kappas … is used in all area".

These claims are questionable in view of the following facts.

− As explained in our letter, in the original paper, the wrong kappa (0.61) between bibliometrics (F) and peer review (P) (labelled "F vs P") in Economics and Statistics, far from being a factual editing error, is exactly the value which is obtained by using the "VQR weights" described in Table 3. These weights were reported in both the VQR report (page 22 of Appendix B of the VQR 2004-2010 Final Report) and in Ancaiani et al. (2015). The same happens for the wrong kappa (0.46) between two reviewers (labelled "P1 vs P2") in Economics and Statistics: rather than being an editing error, it coincides with the value which is obtained by using such "VQR weights". Hereafter we denoted this set of weights as $VQR_{wrong}$; (note that with the available data we cannot replicate results for the third retracted kappa, relative to Agriculture and Veterinary);

---
[3] This is a second problem for the integrity of research records as published in *Research Evaluation.* Indeed the journal did not publish a correction and readers of Ancaiani et al. (2015) will not be able to know immediately that the Table 3 of the paper contains wrong data and that it should be replaced by Table 3 of Benedetto et. Al (2017).

**Table 3.** The VQR$_{wrong}$ weights.

|  | | Informed peer review/P2 | | | |
|---|---|---|---|---|---|
|  | | A | B | C | D |
| **Bibliometrics /P1** | A | 1 | 0,8 | 0,5 | 0 |
|  | B | 0,8 | 1 | 0,8 | 0,5 |
|  | C | 0,5 | 0,8 | 1 | 0,8 |
|  | D | 0 | 0,5 | 0,8 | 1 |

Source: Table 3 of (Alberto Baccini and De Nicolao 2017)

- The corrected kappas ("F e P, VQR weights"; "P1 e P2, VQR weights") for Economics and Statistics in the new Table 3 now coincide with a different set of weights identically labelled as "VQR weights" and used by (Bertocchi et al. 2015). These weights, hereafter denoted as VQR$_{correct}$, are reported in Table 4.

**Table 4.** The VQR$_{correct}$ weights.

|  | | Informed peer review/P2 | | | |
|---|---|---|---|---|---|
|  | | A | B | C | D |
| **Bibliometrics /P1** | A | 1 | 0,8 | 0,5 | 0 |
|  | B | 0,8 | 1 | 0,7 | 0,2 |
|  | C | 0,5 | 0,7 | 1 | 0,5 |
|  | D | 0 | 0,2 | 0,5 | 1 |

Source: Table 4 of (Alberto Baccini and De Nicolao 2017)

- Benedetto et al. retract just three kappas (Economics and Statistics: F vs P and P1 vs P2, and Agriculture and Veterinary: P1 vs P2) and "confirm that the same set of weights … is used in all area". We are therefore left to believe that all the other kappas, both in the VQR official Report (ANVUR 2013) and in (Ancaiani et al. 2015), had already been computed by using the VQR$_{correct}$ weights. This would mean that Ancaiani et al. (and also ANVUR) had computed 17 kappas out of 20 by using the VQR$_{correct}$ weights, never mentioned in their documents; or equivalently that only 3 kappas out of 20 were calculated by using the set of VQR$_{wrong}$ weights, the only ones reported in Ancaiani et al.'s paper (and also, repeatedly, in ANVUR's reports);

- Differently from the kappas of Economics and Statistics, whose inconsistency emerged because their calculation was replicable, no one can say whether 18 kappas out of 20 were calculated by VQR$_{wrong}$, as declared by both ANVUR's report and Ancaiani et al. (2015), or by VQR$_{correct}$ as now claimed by Benedetto et al.. Analogously, in ANVUR's official reports, 43 kappas are published also for the sub-areas (Alberto Baccini and De Nicolao 2016a). While for the 4 sub-areas of economics and statistics we have verified that kappas are computed by using the VQR$_{correct}$ weights, it is impossible to verify what system of weights was used in the other 39 sub-areas. In sum: if raw data are not disclosed, it is impossible to verify the statement of Benedetto et al. according to which a same set of weights is used for all areas (and sub-areas).

4. **Undisclosed averaging procedures.**

Benedetto et al. introduce a central argument not so highlighted in their original paper. After having admitted that the degree of agreement between peer review and bibliometrics is "poor to fair", they state that this agreement "is on average and in most higher than the agreement between two reviewers of the same paper". Here there are two problems, the former already illustrated in (Alberto Baccini and De Nicolao 2016a, 2016b):

- The value of P for each paper is decided by "averaging" two referee scores. This was not done by an automatic algorithm (see page 8 of Appendix B of the VQR 2011-2014 Final Report: "in the VQR 2004-2010 the synthetic evaluation P was instead referred to the final evaluation by the GEV panel and not to the simple arithmetic synthesis of the reviewers scores"). Despite our effort (Alberto Baccini and De Nicolao 2016a, 2016b), we were unable to identify a simple procedure in the official reports explicating how the score was averaged for papers for which the two referees disagreed in their evaluation. However, we documented that many different systems were adopted by different areas. We documented also that for many papers the average score was directly decided by a consensus group formed by two members of the Area panel. These two panellists, at least in economics and statistics, knew not only the bibliometric score of the paper but they knew also that the value of P would be used for evaluating agreement between peer review and bibliometrics. Consider for example a paper having received "A" as bibliometric evaluation; assume also that the score assigned by the first peer (P1) is "A", and the score by the second peer (P2) is "B". What's the average score P used for calculating kappas? If P="A" is assigned, this decision pushes up kappas for the agreement between "F and P", by favouring the result of a better agreement between F and P rather than between P1 and P2. In sum, the now "central argument" of Benedetto et al. depends from decisions about the average P, that were controlled by panellists. Since data about P1, P2 and P are not disclosed, it is impossible to verify how the average P was calculated, and if these calculations induced some bias in the final results of the paper;

- The comparison of the P1 vs P2 agreement with the F vs P one is not really meaningful. If we take the P evaluation as a benchmark, the F vs P agreement should rather be compared with the P1 vs P agreement and the (equivalent) P2 vs P agreement. Given that P is a synthesis between P1 and P2, the agreement between P1 and P (and P2 vs P) will be obviously better than that between P1 and P2. Is the P1 vs P (P2 vs P) agreement also better than the F vs P one? Again, the comparison cannot be done because data are not available.

5. **Official reports contradict the claim of "same protocol in all areas".**

Benedetto et al. wrote: "We confirm that … the same protocol [has] been used in all areas. The peculiarities regarding Area 13 are largely discussed in Bertocchi et al. 2016 and Bertocchi et al (2016)". Benedetto et al. failed to cite that (Alberto Baccini and De Nicolao 2016a, 2016b) documented, by using ANVUR official reports., that protocols used for economics and statistics was radically different from the ones of the other areas, and that this difference may be responsible of the better agreement between peer review and bibliometrics in economics and statistics.

6. **Unfaithful attribution.**

Benedetto et al. attributed to us unfaithfully a claim "that peer review and bibliometrics cannot be used interchangeably". In our letter we limit to comment that on the basis of their data, their conclusion ("results of the analysis relative to the degree of concordance and systematic difference may be considered to validate the general approach of combining peer review and bibliometric methods") appears to be unsound.

**Conclusion.**

The article by Ancaiani et al. (2015), our letter, the reply by Benedetto et al. (2017) and this note contain not only different opinions but also irreconcilable data. We recall that, as mentioned in our letter, we had asked a copy of the raw data without receiving any response. This discussion happens therefore under a condition of asymmetric information that prevents us not only from reproducing results presented in Ancaiani et al. (2015), Benedetto et al. (2017) but even from ascertaining the consistency of their data.

Hopefully, this note will send a warning to the scholarly community interested in research evaluation, bibliometrics and research policy. Several papers, usually authored by ANVUR employees or by scholars serving in ANVUR committees, rely on ANVUR data, and particularly on data from the "ANVUR experiment" about concordance between bibliometrics and peer-review. As long as these data remain not accessible to the scientific community, scholars should be made aware that results of these papers, such as, last among the others, (Jappelli et al. 2017) or (Bonaccorsi et al. 2017), are currently not-reproducible. It is therefore difficult to consider data and conclusions of these articles as sound piece of science at least until authors or ANVUR disclose the data necessary for replication.

In the case at hand, it is hardly understandable why data are retained by ANVUR. In fact, the publication of anonymous data, as listed in the Appendix A of this note, would suffice to clarify at least the following three major issues: (1) establishing the true size of the subsample; (2) calculating statistics on the representativeness of sample and subsample at a sub-area level; (3) verifying how P1-P2 scores were averaged.

A last consideration. In Italy the performance-based research funding system adopted by the government and realized by ANVUR is based on evidence that scholars cannot control or replicate.

**APPENDIX A. Description of the dataset needed for replication**

*# of items (articles of the random sample): 9,199*

For each item:

Area: indication of the area in which the paper is classified

Sub-area: indication of the sub-area (sub-gev) in which is classified

F: bibliometric evaluation

P1: score assigned by peer 1

P2: score assigned by peer 2

P: final peer-review average score